\documentstyle[aps,version2]{revtex}
\begin{document}
\begin{title}
Long range Coulomb forces and
the behaviour of the chemical potential of electrons
in metals at a second order phase transition.
\end{title}
\author{D. van der Marel, D. I. Khomskii}
\begin{instit}
Laboratory of Solid State Physics, Materials Science Center,
University of Groningen, Nijenborgh 4, 9747 AG Groningen, The
Netherlands
\end{instit}
\moreauthors{G. M. Eliashberg}
\begin{instit}
L. D. Landau Institute for Theoretical Physics, Chernogolovka, 142432 Russia
\end{instit}

\receipt{}
\begin{abstract}
We give a general thermodynamic analyzis of the behaviour of the chemical
potential of electrons in metals
at a second order phase transition, including in our analysis the
effect of long range Coulomb forces. It is shown, that this chemical
potential can have a kink at T$_c$, both for fixed sample volume and
fixed external pressure. The Coulomb term
transfers the changes in chemical potential of the electrons
into an experimentally observable shift of the surface potential if the
sample is electrically connected to a ground potential.
\vspace{2 \baselineskip}
\end{abstract}
\widetext

\section{Introduction}

The behaviour of the chemical potential ($\mu^e$) of electrons
at the superconducting
phase transition was discussed recently in Ref. \cite{mu1,mu2,mu3}, and
corresponding measurements were carried out in Ref. \cite{mu4}. It was
shown theoretically, that in the BCS model $\mu^e$ is given by the expression
$\mu^e=\mu_0^e-\frac{\Delta^2}{4\mu_0}$, or, if the density of states near
$E_F$ is energy dependent $\mu^e=\mu_0^e(1-\frac{d\rho/dE}{\rho}\Delta(T)^2)$,
so that
it has a kink at $T_c$. General consideration gives for the kink the
expression
 \begin{equation}
   \frac {\Delta \left( d\mu^e/dT \right)_V}{\Delta C_V } =
    \frac {d \ln T_c}{d N_e}
 \label{eq:1}
 \end{equation}
\noindent
where $N_e$ is the number of electrons and $\Delta C_V$ is the jump
in specific heat.
This conclusion was confirmed experimentally\cite{mu4},
where it was shown that $\mu^e$ in YBa$_2$Cu$_3$O$_{7-x}$ does indeed have
a kink at T$_c$. This result is of an essentially thermodynamic nature and is
actually independent of the specific nature of the phase transition, {\em i.e.}
it is valid for other second order phase transitions as well, not only
for superconductivity. \\
However, in Ref. \cite{mu3} relation Eq.\ref{eq:1} was derived for a
system at fixed volume, whereas the standard experiments are carried out
at constant ambient pressure. For a system at constant presssure
we should use
the Landau expansion not for the Helmholz free energy $F(V,T,N,\psi)$, but
the  Gibbs free energy $G(p,N,T,\psi)$, where $\psi$ is an order parameter.
Hence the well-known treatment \cite{lanlif} of phase
transitions seems at first sight to rule out the possibility of
obtaining a kink in $\mu$: As the chemical potential is just this
Gibbs free energy per particle, $G=N\mu$, a kink in $\mu$ is
the hall-mark of a {\em first} order phase transition\cite{lanlif},
hence it seems paradoxal that a kink in the chemical potential
was observed experimentally in a second order superconducting
phase transition.
\\
The situation, however, is not so simple, and the
solution goes down to the definition of the chemical potential. The
point is that, when dealing with metals, one usually discusses not the total
chemical potential $\mu$
determined as a Gibbs free energy per elementary cell or per mole of the
substance, but the chemical potential of the electrons $\mu^e$ (which at
$T=0$ is equal to the Fermi energy). One may say
that the total chemical potential $\mu$ is the Gibbs free energy for adding to
the system an extra unit cell, with all its nuclei and electrons, whereas
$\mu^e$ is the change in $G$ when we change only the concentration of
electrons. Thus, generally speaking, the situation here is similar to the
situation in a mixtures, solutions or a complex chemical
compound\cite{lanlif}. In that case we have to introduce chemical
potentials for each component
\begin{equation}
G=\sum_i\mu_i N_i
\end{equation}
\noindent
and whereas $G$ itself, and the corresponding total chemical potential
$\mu=\partial G/\partial N = G/N$ (where $N$ is the number of formula units
for fixed concentration of components) should have no kink at $T_c$ (this is
actually the definition of a second order phase transition), {\em partial}
chemical potentials $\mu_i$ may well behave differently, as has been
discussed for the high $T_c$ ceramics by Burns\cite{burns}. The situation with
$\mu^e$ is still somewhat more complicated
because, if we want to discuss it separately, {\em i.e.} if we want to look at
the changes in the system when we change the electron concentration, we have
also to take into account long range Coulomb forces, which usually guarantee
electro-neutrality of the system. We will show below that, all these factors
taken care of, the chemical potential of electrons $\mu^e$ as measured {\em
e.g.}
in Ref\cite{mu4} should indeed have a kink at $T_c$, even if in the
experiment not the volume but the pressure is fixed externally.

\section{Single component system.}

First of all we derive Eq. \ref{eq:1} using a slightly different method
than that used in \cite{mu3}, and show that indeed even for a one-component
system the chemical potential at constant volume has a kink at the second
order phase transition, in contrast to the situation at fixed pressure.
At fixed pressure the Gibbs free energy $G(p,T,N)=N\mu(p,T)$, and
$\mu(p,T)$ as well as the first derivatives of $\mu$ are continuous at the
transition. The temperature derivative of $\mu$ at fixed pressure follows
by making the transformation of variables $\mu(V,T)=\mu(p(V,T),T)$,
so that
\begin{equation}
\left(\frac{d\mu}{dT}\right)_V=
\left(\frac{d\mu}{dT}\right)_p+
\left(\frac{d\mu}{dp}\right)_T\left(\frac{dp}{dT}\right)_V
\end{equation}
We furthermore identify $\left(d\mu/dp\right)_T$ as $V/N$.
Only the second term of this expression gives rise to a discontinuity
at the phase transition, for which we can use one of
the Ehrenfest relations\cite{lanlif}
\begin{equation}
\frac{\Delta\left( dp/dT \right)_V}{\Delta C_V} = - \frac{d\ln T_c}{dN}
\label{eq:5ssb}
\end{equation}
As for a single component system $NdT_c/dN=-VdT_c/dV$, we
can now rewrite the Ehrenfest relation in the form
\begin{equation}
   \frac {\Delta \left( d\mu/dT \right)_V}{\Delta C_V } =
    \frac {d \ln T_c}{d N}
\end{equation}
which is identical to Eq.\ref{eq:1}. This treatment resolves the apparent
contradiction between microscopic treatments, which give a finite jump
in $d\mu/dT$ at T$_c$, and general arguments according to which $\mu$
has to be smooth at a second order phase transition: As the microscopic
treatments are carried out for a given electron concentration, and
neglecting temperature dependencies of the volume, if the system
is held at a constant external pressure. As we will show in the next
section, the partial chemical potential of electrons may have a kink,
even if such thermal changes of the volume are taken into
account. Thus it is instructive to look for its behaviour first at
fixed volume, such as is usually obtained from a microscopic description,
and than make the Legendre transformation of the free energy
$G(p,N,T)=F(V,N,T)+pV$, to discuss the effects of fixing external
pressure instead of volume. Important here is the role of the long
range Coulomb interaction
\section{Fixed volume}
The thermodynamical
state of the solid as follows from these conditions is described by
the Helmholtz free energy $F(V,T,N,N_e)$, where
$N$ is the number of elementary cells, and $N_e$ is the number of electrons.
Equivalently we can use the electron concentration $x=N_e/N$, so that we
can write the free energy in the form
$F(V,T,N,x)$. The Gibbs free energy per unit cell is
$\mu(V)=\left(\frac{\partial F}{\partial N}\right)_{V,T,x}$ and the chemical
potential of the electrons is
$\mu^e(V)=N^{-1}\left(\frac{\partial F}{\partial x}\right)_{V,T,N}$. If the
particles would be neutral, the free energy would depend on the volume as
$F = V f(n,x,T)$ where $n=N/V$. However if we want to
discuss phenomena where the number of electrons is allowed to change,
we should also add the
Coulomb (charging) term, which has a different dependence on the size
of the system. We use for the corresponding term the simple expression
$\delta F_C = \frac{e^2 N^2 (x-x_0)^2}{2 V^{1/3}}$ (Coulomb energy of a
charged sphere of radius $R=V^{1/3}$ and charge $eN(x-x_0)$ with $x_0$ the
equilibrium number of electrons per unit cell). We will see later,
that the precise shape of the sample, and therefore the exact value of
the prefactor in $\delta F_C$ is unimportant. Hence the Helmholtz free energy
is
\begin{equation}
 \begin{array}{lll}
   F(V,N,x,T) &=& Vf(n,x,T) + \frac{1}{2}e^2(x-x_0)^2 N^2 V^{-1/3} \\
 \end{array}
\label{eq:6}
\end{equation}
\noindent
 For systems where the external volume is fixed, the chemical
 potential of the electrons is
\begin{equation}
\begin{array}{lll}
\mu^e(n,x,T,\phi) &=& \left.\frac{1}{N} \frac{d F}{d x}\right|_{V,N,T} =
      \frac{1}{n} \frac{\partial f}{\partial x}  + e\phi \\
\end{array}
\label{eq:7}
\end{equation}

\noindent
 where in the last line we introduced the charging potential
 $\phi=e (x-x_0) n^{1/3} N^{2/3}$.
 If the sample is externally grounded, the electrons are in equilibrium with
 the ground potential which fixes $\mu^e$ at the
 value of the external bath. The electrons can now flow freely in and out of
the
 sample, so that the number of electrons follows from solving the above
 expression for $x$. From this we see that the charge per unit
 cell $x-x_0$ is proportional to $N^{-2/3}$. Note that the total chemical
 potential has a different form:
\begin{equation}
\begin{array}{lll}
\mu(n,x,T) &=& \left.\frac{d F}{d N}\right|_{V,x,T} =
       \frac{\partial f}{\partial n}  +
        \phi^2 n^{-1/3} N^{-2/3} \\
\end{array}
\label{eq:7sb}
\end{equation}
\noindent
 We see that, due to the fact that $\phi$ is finite, the
 last (Coulomb) term vanishes in the thermodynamic limit. Clearly, as
 different partial derivatives are taken of the functional $f(n,x,T)$,
 also a different behaviour at the phase transition
 occurs for $\mu$ and $\mu^e$. Both of them, however, would
 have a kink at T$_c$: Using the continuity of entropy
 along the curve $T_c(x)$ in the former,
 and $T_c(n)$ in the latter case (as in Ref. \cite{mu3}), one obtains
 Eq. \ref{eq:1} for the kink in $d\mu^e/dT$, while
 \begin{equation}
     \Delta \left(\frac{d\mu}{dT}\right)_V=
     N^{-1} \Delta C_V \frac {d \ln T_c}{d \ln n}
 \end{equation}
 Hence the kink in $d\mu/dT$ is proportional to the
 derivative of $T_c$ with respect to the density of {\em unit cells}, which
 may also include a possible redistribution of electrons
 between reservoirs upon changing the lattice constant. On the other
 hand  $\Delta \left( d\mu^e/dT \right)$ is proportional to the
 $dT_c/dx$, where $x$ is the number of electrons per unit cell. As
 was treated in Ref. \cite{mu2}, if the solid contains two
 charge reservoirs, one of which is 'active' in
 the phase transition, a reduction of
 $\Delta \left( d\mu^e/dT \right)$
 occurs if a charge re-distribution between the reservoirs takes place
 at the phase transition. This
 also follows from the present thermodynamic analysis: $dT_c/dx$ will be
 reduced, if part of the electrons moves into an 'inactive' reservoir
 upon varying $x$.
 \section{Fixed pressure}
 First of all we again obtain the general thermodynamic expression for
 the behaviour of $\mu^e$ and
 $\left(d\mu^e/d T\right)_p$
 initially ignoring charging effects. Following \cite{lanlif2} we
 differentiate the expression
 $\Delta\left(\partial G/\partial T\right)=0$
 (continuity of entropy) along the curve $T_c(x)$ and find
 \begin{equation}
 0=\frac{d}{dx}\Delta\left(\frac{\partial G}{\partial T}\right)=
 \left[ \frac{dT}{dx}
       \Delta\left(\frac{\partial^2 G}{\partial T^2}\right)
    +  \Delta\left(\frac{\partial^2 G}{\partial T\partial x}\right)
 \right]_{T=T_c}
 \label{eq:9}
 \end{equation}
so that
 \begin{equation}
   \Delta\left(\frac{d\mu^e}{dT} \right)_p = \frac{\Delta C_p}{N}
    \frac {d \ln T_c}{d x}
 \label{eq:10}
 \end{equation}
 Thus we see, that indeed if we could treat the electron concentration
 $x$ as an independent parameter, the chemical potential of electrons $\mu^e$
 would have a kink at the second order phase transition, even at fixed
 pressure. Expression \ref{eq:10} has the same structure as Eq. \ref{eq:1},
 with the natural change of variables.
 \\
 We can take into account charging effects, by going from $F(V,N,x,T)$ to
 the Gibbs free energy $G=F+pV$, where $p=-\partial F/\partial V$,
 and calculate $\mu^e(p,N,x,T)\equiv\partial G/\partial N_e$ and
 $\mu(p,N,x,T)\equiv\partial G/\partial N_e$. As a result of the
 Legendre transformation, these have the same functional form as
 Eqs. \ref{eq:7} and \ref{eq:7sb}. However, these
 expressions are in this case a function of
 $p$, as the dependence of the density $n$ on the external pressure
 has to be solved from
 \begin{equation}
  \begin{array}{lll}
    p(n,x,T) &=& - f(n,x,T) + n \frac{\partial f}{\partial n}
        + \frac{1}{6} \phi^2 n^{2/3} N^{-2/3} \\
      \label{pn}
  \end{array}
 \end{equation}
 As before $\phi$, defined as $e (x-x_0) n^{1/3} N^{2/3}$, is the
 charging potential. Note that the Coulomb term introduces an
 additional dependence on the total
 number of unit cells $N$, due to its long range nature. Hence
 in principle we should consider $n(p,T,x,N)$. However, this contribution
 vanishes for $N \rightarrow \infty$.
 \begin{equation}
 \begin{array}{lll}
 \mu^e(p,x,T,\phi) &=& \frac{1}{n} \frac{\partial f}{\partial x}
 + e \phi \\
 \mu(p,x,T) &=& \frac{\partial f}{\partial n} +
 \phi^2 n^{-1/3} N^{-2/3} \\
 \label{mupt1}
 \end{array}
 \end{equation}
\noindent
 As $\mu$ has to be
 evaluated at the minimum with respect to the order parameter, it is
 automatically ensured that it has no kink at the phase transition,
 provided that a Landau expansion can be made. We also notice, that
 the charging effects do not affect our expression for the kink
 in $\mu^3$ (Eq. \ref{eq:10}), as the charging term is of vanishing
 order for $N \rightarrow \infty$: If the sample is electrically
 isolated $N$ and $x$ are fixed. As $\phi=e(x-x_0)n^{1/3}N^{2/3}$,
 the only thermal variations enter through the volume changes (changes
 of $n$). Hence, if the sample is electrically charged, there are
 temperature dependent changes to $\mu^e$ due to changes in lattice
 constant at fixed external pressure, that should be taken into account.
 If, on the other hand the sample is electrically grounded, $\mu^e$ is
 fixed externally. In this case $e\phi$ is exactly
 equal to $\mu^e$ (but with opposite sign) of the electrically isolated
 uncharged sample, and can be measured experimentally.

\section{Example}

For a solid undergoing a
second order phase transition, we
may write $f(n,x,T,\psi)=f_0+f_{\psi}$, where for
$f_0$ and $f_\psi$ we make the following free energy expansion

 \begin{equation}
 \begin{array}{lll}
   F_0(N,V,T,\psi) &=& (\mu_0+\mu_{\psi}) N +
\frac{1}{2}B_0V_0(\frac{V}{V_0}-1)^2
   +c(V-V_0)|\psi|^2 + \frac{1}{2}\phi^2V^{1/3}\\
   \mu_{\psi}(n,T,\psi) &=& a(T-T_c)|\psi|^2+\frac{b}{2}|\psi|^4  \\
 \end{array}
 \end{equation}

\noindent
It has to be understood here, that the parameters $a$, $b$, and $T_c$ are
independent of the volume $V$, and coupling between the strain field
and the superconducting order parameter is introduced in an ad-hoc manner
through the constant $c$.
The expression for the pressure as occurs in Eq. \ref{pn} is

\begin{equation}
p=-B_0\left[\frac{V}{V_0}-1\right]-c|\psi|^2 + \frac{1}{6}\phi^2V^{-2/3}
\end{equation}

from which we solve the volume $V=V_0\left\{1-(p+c|\psi|^2)/B_0\right\}$. We
are now ready to calculate the Gibbs free energy, and from it, with the
definition of $\mu$, the chemical potential

\begin{equation}
\mu(p,T,\psi)=\mu_0+\frac{p}{n_0}\left[1-\frac{p}{2B_0}\right]
 +\left[a(T-T_c)-\frac{c}{n_0}\frac{p}{B_0}\right]|\psi|^2
 +\frac{1}{2}\left[b-\frac{c^2}{n_0B_0}\right]|\psi|^4
 + \frac{2}{3}\phi^2 n^{-1/3}N^{-2/3}
 \end{equation}

Here $n_0\equiv N/V_0$ represents the molar density of the solid
if the order parameter is zero.
So in the first place we notice, that in spite of the coupling to the lattice,
the transition remains of second order. Only if the coupling constant
$c$ exceeds the critical value $(bn_0B_0)^{1/2}$, it
becomes of first order. In the second place we notice that, in the case
where the transition is still of second order, $T_c$ is shifted
with an amount and direction which depends on $c$ and the value of
the externallly applied pressure. The Gibbs free energy per particle
calculated at the minimum with respect to $\psi|^2$ is

\begin{equation}
\begin{array}{lll}
\mu(p,T)&=&\mu_0+\frac{p}{n_0}\left[1-\frac{p}{2B_0}\right]
-\frac{a^2(T-\tilde{T}_c)^2}{2\tilde{b}}
+\frac{2}{3}n^{-1/3}N^{-2/3}\\
      \tilde{T}_c(p)&=&T_c+p\frac{c}{n_0B_0a}\\
      \tilde{b}&=&b-\frac{c^2}{n_0B_0}
\end{array}
\end{equation}
\noindent

Clearly the total chemical potential of the system has a
smooth behaviour at the phase transition without a kink, unless the
coupling to the lattice is sufficiently strong to make the transition
first order. \\
Let us now consider the electronic sub-system.
As we are considering a solid, the number of unit cells, $N$,
is externally fixed, and the corresponding
chemical potential need not be in equilibrium with
an external bath. The situation is completely different for the
electrons however, as these can move in and out of the solid. Using
again the definition $x=N_e/N$ for the number of electrons per unit
cell, we can calculate the corresponding chemical potential by
differentiating the total Gibbs free energy ($\mu N$, where $\mu$ is the
chemical potential as calculated above) with respect to $N_e$. As we
have seen in the preceeding section, the long range Coulomb forces can
be included, and the correct expression becomes
$\mu^e(p) = \left. \frac{\partial \mu}{\partial x} \right|_{p,T} + e\phi$,
where we have to assume now that $\mu$ not only depends on $p$ and
$T$, but also on $x$. Indeed most of the properties of a solid depend
strongly on the number of charge carriers per unit cell, for example by
influencing the strength of the chemical bond between neighbouring
unit cells, or by having an effect on the superconducting transition
temperature. For our discussion the latter dependence is the most
important one, as we are interested in the behaviour near the superconducting
transition. Let us indeed assume that $\mu_{\psi}$
is derived from a microscopic theory, which also predicts that
$dT_c/dx\neq 0$. If the electronic subsystem is brought into equilibrium
with an external bath by connecting it with a current lead, the charge on
the sample is such, that $e\phi$ compensates for the difference. Hence
$\mu^e(p)=\mu^{ext}$ and the voltage on the sample is

\begin{equation}
e\phi = \mu^{ext} - \mu_n -
  \frac{a^2(T-\tilde{T}_c)}{\tilde{b}}  \frac{d\tilde{T}_c}{dx}
 \label{eq:ephi}
\end{equation}

This voltage has a kink at $\tilde{T}_c$, which
can be determined experimentally by measuring the workfunction of the sample.
If the lattice is sufficiently soft
($B_0 \le c^2/n_0b$) the transition becomes of first order. The above
expression diverges at the point where $B_0 = c^2/n_0b$ and
is no longer valid in this limit.
If the material is hard ($B_0 \gg c^2/n_0b$) and the external pressure
small ($ p \ll n_0aB_0/c$) Eq. \ref{eq:ephi} is just the result which
we already obtained at fixed sample volume.

\section{Conclusions}
The seemingly paradoxal result of microscopic theories of superconductivity,
that the chemical potential may have a kink at a {\em second order} phase
transition, is resolved.
There are two ingredients in the resolution of the paradox. The first is, that
the microscopic treatment is always carried out
at a given concentration of electrons,
{\em i.e.} at fixed volume or density. In contrast to fixed pressure, there is
absolutely no general rule forbidding the kink in $\mu$ at fixed volume;
general thermodynamic results\cite{mu3} confirm that.
More interesting is the second part of the story. The chemical potential
of the electrons is in general different from the total chemical potential
and thus can have a kink, even at fixed pressure. This stems from the fact
that electrons are free to move in and out of a solid, thus maintaining
electrical equilibrium with the environment, whereas the ions
in a solid are immobile. As a result the chemical potential of electrons
and the Gibbs free energy per mole of the ions
have different dependencies on temperature, pressure etc., and
are only coupled through the long range Coulomb forces.
Using scaling arguments the Coulomb charging energy is shown
to be of vanishing order in
the Gibbs free energy per unit cell, the density of the solid, T$_c$ and
the order parameter. At the same time the long range Coulomb forces on
the one hand keep the charge carrier density fixed, while on the other hand
they transform changes of the chemical potential of the electrons into
equally large and measurable changes of the workfunction, if the solid
is in electrical equilibrium with its environment.

\section{Acknowledgements}
This investigation was supported by the Netherlands Foundation for
Fundamental Research on Matter (FOM).


\begin{references}
\bibitem{Tx} In principle $T_c$, $a$, and $b$  can depend on $n$, $x$, {\em
 as well as} $x_0$ which parametrizes the chemical composition if the sample.
 However, only the explicit dependence of $T_c$ on $x$, contributes to the
 kink in $\mu^e$ at $T_c$.
\end{references}
\end{document}